\documentclass[pra,10pt,twocolumn,superscriptaddress,floatfix,showpacs,groupedaddress]{revtex4-1}
\usepackage[utf8]{inputenc}  
\usepackage[T1]{fontenc}     
\usepackage[british]{babel}  
\usepackage[sc,osf]{mathpazo}
\usepackage[scaled=0.86]{berasans}  
\usepackage[colorlinks=true, citecolor=blue, urlcolor=blue]{hyperref}  
\usepackage{graphicx} 
\usepackage[babel]{microtype}  
\usepackage{amsmath,amssymb,amsthm,bm,amsfonts,mathrsfs,bbm} 

\usepackage{xspace}  
\usepackage{pgfplots}

\begin{document}
\title{Limited preparation contextuality in quantum theory and its relation to the Cirel'son bound}

\author{Manik Banik}

\author{Some Sankar Bhattacharya}
\author{Amit Mukherjee}
\author{Arup Roy}
\affiliation{Physics and Applied Mathematics Unit, Indian Statistical Institute, 203 B. T. Road, Kolkata 700108, India.}

\author{Andris Ambainis}
\author{Ashutosh Rai}
\affiliation{Faculty of Computing, University of Latvia, Raina bulv. 19, Riga, LV-1586, Latvia.}

\begin{abstract}
Kochen-Specker (KS) theorem lies at the heart of the foundations of quantum mechanics. It establishes impossibility of explaining predictions of quantum theory by any noncontextual ontological model. Spekkens generalized the notion of KS contextuality in [\href{http://link.aps.org/doi/10.1103/PhysRevA.71.052108}{Phys. Rev. A {\bf 71}, 052108 (2005)}] for arbitrary experimental procedures (preparation, measurement, and transformation procedure). Interestingly, later on it was shown that preparation contextuality powers parity-oblivious multiplexing [\href{http://journals.aps.org/prl/abstract/10.1103/PhysRevLett.102.010401}{Phys. Rev. Lett. {\bf 102}, 010401 (2009)}], a two party information theoretic game. Thus, using resources of a given operational theory, the maximum success probability achievable in such a game suffices as a \emph{bona-fide} measure of preparation contextuality for the underlying theory. In this work we show that preparation contextuality in quantum theory is more restricted compared to a general operational theory known as \emph{box world}. Moreover, we find that this limitation of quantum theory implies the quantitative bound on quantum nonlocality as depicted by the Cirel'son bound.       
\end{abstract}

\maketitle
Quantum mechanics (QM) departs fundamentally from the well known \emph{local-realistic} world view of classical physics. This stark contrast of quantum theory with classical physics was illuminated by J. S. Bell \cite{Bell}. Since the Bell's seminal work, nonlocality remains at the center of quantum foundational research \cite{Mermin,Brunner}. More recently quantum nonlocality has been also established as a key resource for device independent information technology \cite{Brunner,Scarani}. Quantum nonlocality does not contradict the relativistic causality principle, however, QM is not the only possible theory that exhibits nonlocality along with satisfying the no-signaling principle; there can be non-quantum no-signaling correlations exhibiting nonlocality. One extreme example of such a correlation (more nonlocal than QM) was first constructed by Popescu and Rohrlich (PR) \cite{Popescu}. Whereas PR correlation violates the Bell-Clauser-Horne-Shimony-Holt (Bell-CHSH) \cite{Bell,CHSH} inequality by algebraic maximum, the optimal Bell-CHSH violation in quantum theory is restricted by the Cirel'son bound \cite{Cirelson}. In this work, we show that Cirel'son limit on nonlocal behavior of quantum theory can be explained from its another very interesting feature, namely restricted \emph{preparation contextuality}.  

Nearly at the same time of Bell's result, Kochen and Specker proved another important no-go theorem showing that predictions of \emph{sharp}  (projective) measurements in QM can not be reproduced by any non-contextual ontological model \cite{KS}. Unlike Bell-nonlocality, structure of QM is implicit in the definition of KS contextuality. However, recently the idea of KS contextuality has been generalized, by Spekkens \cite{Spek05}, for arbitrary operational theories rather than just quantum theory and for arbitrary experimental procedures rather than just sharp measurements. It was then shown that mixed preparations (density matrices) in quantum theory exhibit preparation contextuality \cite{Spek05,Banik}. Interestingly, invoking another non-classical concept called \emph{steering} \cite{Schrodinger,Wiseman} along with this new idea of preparation contextuality one can establish nonlocality of QM without using any Bell type inequalities; it has been shown that nonlocality of some hidden variable models, underlying QM, directly follows from the steerability of bipartite pure entangled states and the preparation contextuality of mixed states \cite{Banik,Harrigan,Leifer}.

The traditional definition of contexuality address to only the contexts of projective measurements, which has been studied in much depth \cite{Peres, Grudka}. However, generalized notion of contextuality developed by Spekkens define three different types of contexts: measurement (generalized), preparation, and transformation contexts \cite{Spek05}. Interest in studying contextuality in this general framework is relatively new and growing \cite{Banik, Harrigan, Leifer, Spek09, Ravi, Mazurek, Pusey}. This generalized approach has lead to designing more robust experimental tests of contextuality \cite{Ravi,Mazurek, Pusey}; these recent results are very promising given that previously requirements for testing contextuality in experiments has been a topic of much controversy ( for a more discussion see the ref. \cite{Pusey} and relevant references therein). 

Our work here is build upon the notion of preparation contexuality which address the impossibility of representing two equivalent preparation procedures, of an operational theory, equivalently in some ontological model. More precisely, suppose we have two equivalent operational preparations $P,P'$, i.e., the outcome probabilities $p(k|P,T,M)=p(k|P',T,M)~\forall$ outcomes $k$, transformations $T$, and measurements $M$. Then a hidden variable (ontic) model, which reproduces $p(k|P,T,M)$ by averaging over the ontic states $\lambda$ is preparation noncontextual if $\forall M,T:p(k|P,T,M)=p(k|P',T,M)\Rightarrow p(\lambda|P)=p(\lambda|P')$, where $p(\lambda|P)$ and $p(\lambda|P')$ represent respective distributions over the ontic states followed by operational preparations $P$ and $P'$\cite{Spek05}.

Preparation contextuality has operational usefulness as it powers \emph{parity-oblivious multiplexing} (POM), a two-party secure computation task \cite{Spek09}; in this work, Spekkens \emph{et al.} derived a `noncontextuality inequality' which place an upper bound on any operational theory that admits a preparation noncontextual ontological model. Further the authors showed that the success rate of the POM game played with only classical resources is restricted by the same inequality. Thus, Spekkens and coauthors concluded that any operational theory is preparation contextual if it can beat the classical bound in a POM task.  


It turns out that, in performing a POM task, certain quantum resources can do better than any classical resource, thus proving that QM is a preparation contextual theory. Interestingly, in this work we show that for performing a POM task, there exists an operational (toy) theory, namely \emph{box world} \cite{Barrett,Janotta}, which can do better than QM. Thus, though QM is preparation contextual, the amount of preparation contextuality in QM is constrained compared to the box world. Furthermore, we show that restricted preparation contextuality of quantum theory leads to its limited nonlocal behavior as depicted in the Cirel'son bound. Therefore, our result brings the qualitative connection between preparation contextuality and nonlocality explored in \cite{Harrigan,Leifer,Banik} to a quantitative footing.

\emph{Parity-oblivious multiplexing}: It is a variant of the well studied information-processing task called \emph{random access code} (RAC) \cite{Wiesner,Ambainis99,Ambainis02}. Suppose a n-bit string $x$, chosen uniformly at random from $\{0,1\}^n$, is given to Alice. An integer $y$, chosen uniformly at random from $\{1,2...,n\}$, is given to Bob, now task for Bob is to guess the $y^{th}$ bit of Alice's input. Let us denote Bob's guess as $\beta_y$. In the POM game Alice and Bob collaborate to optimize the guessing probability $p(\beta_y=y^{th} \mbox{~bit~of~Alice})$. Alice can send to Bob any information which encodes her input. However, there is a cryptographic constraint: no information about any parity of $x$ can be transmitted to Bob. More specifically, letting $s\in\mbox{Par}$ where  $\mbox{Par}\equiv\{r|r\in\{0,1\}^n,\sum_ir_i\ge2\}$ is the set of $n$-bit strings with at least $2$ bits that are $1$, no information about $x.s=\oplus_ix_is_i$ (termed the $s$-parity) for any such $s$ can be transmitted to Bob (here $\oplus$ denotes sum modulo $2$). 

The main result of Spekkens \emph{et al.}\cite{Spek09} can be now stated more precisely: \emph{ for $n$-bit POM game played with states (resources) from a preparation non-contextual theory, the average success probability is bounded as follows:} $$p_{NC}(\beta_y=y^{th} \mbox{bit~of~Alice})\le\frac{1}{2}(1+ \frac{1}{n}).$$ Motivated by this result, in our work, we define the maximum success probability in a POM task in an operational theory as a \emph{bona-fide} measure to quantify the strength of preparation contextuality of the concerned theory. The approach we adopt here is similar to defining the strength of nonlocality of correlations as the amount of Bell-CHSH violation (or maximum success probability in a Bell-CHSH game). In remaining of this paper, we focus on $2$-bit POM task. For the 2-bit POM game we adopt two different schemes: (1) encoding-decoding scheme and (2) correlation assisted scheme.

\emph{(1) Encoding-decoding scheme}: Alice and Bob can perform POM task by using resources of an general operational theory. In an operational theory, the primitives of description are preparations and measurements (for simplicity, here we do not consider dynamics/transformation of the system) \cite{Barrett,Janotta,Busch,Barnum,Pfister,Barnum,Stevens,Namioka}. The theory simply provides an algorithm for calculating the probability $p(k|P,M)$ of an outcome $k$ of measurement $M$ given a preparation (state) $P$. The collection of all states, that the system can be prepared in, forms a compact and convex subset $\Omega$ of a finite dimensional vector space $V$. Results of a measurement on any state $\omega$ of the theory is described by an \emph{effect} $e:\Omega\rightarrow[0,1]$, which is a map such that $e(\omega)$ is the probability of obtaining the outcome $e$. There is an \emph{unit effect} $u$ such that $u(\omega)=1 \forall \omega \in \Omega$. Any measurement can now be expressed as some set of effects $\{e_i\}$ such that $\sum_i e_i=u$. 

Alice depending on the input string $x\in\{00,01,10,11\}$, given to her uniformly at random, implements a preparation procedure $P_x\in\Omega_A$ in an operational theory $\mathcal{T}$ and sends the encoded particle to Bob. For each integer $y\in\{1,2\}$, Bob implements a binary-outcome measurement $M_y$, and reports the outcome as his output. The \emph{average} probability of winning is given by:
\begin{eqnarray}
p_{\mathcal{T}}(\beta_y=y^{th} \mbox{bit~of~Alice})\equiv p(\beta_y=x_y)~~~~~~~~~\nonumber\\
=\frac{1}{8}\sum_{y=1}^2\sum_{x\in\{00,01,10,11\}}p(\beta_y=x_y|P_x,M_y).\label{success}
\end{eqnarray}
The optimal success probability in an operational theory is $p^{opt}_{\mathcal{T}}(\beta_y=x_y):=\max_{P_x,M_y}p(\beta_y=x_y)$, where optimization is performed over all possible encodings and decodings procedure allowed in the theory $\mathcal{T}$. Of course the encoding and decoding must satisfy the parity-oblivious constraint expressed here as:
\begin{eqnarray}
p(P_{00}|k,M)+p(P_{11}|k,M)= p(P_{01}|k,M)\nonumber\\
+p(P_{10}|k,M);~~\forall~ M\in\mathcal{M},~and ~\forall~ k.
\end{eqnarray}

First we show that optimal success probability in box world is strictly greater than that of quantum theory, i.e., $p^{opt}_{box}(\beta_y=x_y)>p^{opt}_Q(\beta_y=x_y)$. To prove this result we first consider the quantum case, and then the box world.
 
\emph{(a) Quantum theory}: Alice encodes her $2$ bits into the four pure qubits with Bloch vectors $\{(\pm 1,0,0),~(0,0,\pm 1)\}$ equally distributed on the equatorial $X-Z$ plane of the Bloch sphere; as shown in Fig.(\ref{fig1}). Bob performs the measurement $(\sigma_{x}+\sigma_{z})/\sqrt{2}$ if he wishes to learn the first bit, and the measurement $(\sigma_{x}-\sigma_{z})/\sqrt{2}$ if he wishes to learn the second. He guesses the bit value $0$ upon obtaining the positive outcome, otherwise he guesses the bit value $1$. In all cases, the guessed value is correct with probability $\frac{1}{2}(1+\frac{1}{\sqrt{2}})$, which results the average success probability $p_Q(\beta_y=x_y)=\frac{1}{2}(1+\frac{1}{\sqrt{2}})>\frac{2}{3}=p^{opt}_{NC}(\beta_y=x_y)$. Since the parity $0$ and parity $1$ mixtures in this protocol are represented by the same density operator, no information about the parity can be obtained by any quantum measurement. Interestingly, the qubit protocol just described turns out to be quantum optimal.

{\bf Proposition-$1$:} \emph{In a 2-bit POM game, optimum average success probability over all quantum encoding-decoding schemes is: $p^{opt}_Q(\beta_y=x_y)=\frac{1}{2}(1+\frac{1}{\sqrt{2}})$.}

\emph{Proof:} Alice prepares and encodes as $\{ij\longmapsto \rho_{ij}: i,j\in \{0,1\} \}$, where $\rho_{ij}$ are state operators acting on $\mathbb{C}^d$; she can always find an appropriate pure state $|\Psi_{12}\rangle \in \mathbb{C}^d\otimes \mathbb{C}^d$ and projectors $P_{A},P_{A'}$ such that: $\frac{1}{2}\rho_{00}=\mbox{tr}_1 \{(P_{A}\otimes I)|\Psi_{12}\rangle\}$, $\frac{1}{2}\rho_{11}=\mbox{tr}_1 \{((I-P_{A})\otimes I)|\Psi_{12}\rangle\}$, $\frac{1}{2}\rho_{01}=\mbox{tr}_1 \{(P_{A'}\otimes I)|\Psi_{12}\rangle\}$, $\frac{1}{2}\rho_{10}=\mbox{tr}_1 \{((I-P_{A'})\otimes I)|\Psi_{12}\rangle\}$. Alice performs one of the following projective measurements: (i) $P_{A}\otimes I + (I-P_{A})\otimes I =I\otimes I$, (ii) $P_{A'}\otimes I + (I-P_{A'})\otimes I =I\otimes I$,
on part-$1$ of $|\Psi_{12}\rangle$ and depending on the measurement result she sends part-$2$ to Bob (or discard it). Measurement (i) and (ii) respectively produces two decompositions $\frac{1}{2}\rho_{00}+\frac{1}{2}\rho_{11}$ and $\frac{1}{2}\rho_{01}+\frac{1}{2}\rho_{10}$ for part-$2$ of $|\Psi_{12}\rangle$. Alice in this way prepares and send $\rho_{ij}$ to Bob; assured that parity obliviousness condition is always satisfied.
 
Bob on receiving part-$2$, performs a two outcome projective measurement $\{P_{B},(I-P_{B})\}$ ($\{P_{B'},(I-P_{B'})\}$), if he is asked to guess Alice's first (second) bit, and answers $0(1)$ when measurement outcome is $+1(-1)$. Here $P_{A},P_{A'},P_{B},P_{B'}$ are projectors acting on $\mathbb{C}^d$. Due to no constraint on the dimension of the Hilbert space, Neumark’s theorem allows us to consider only projective measurements, without loss of generality. Substituting $\rho_{ij}$ in terms of $|\Psi_{12}\rangle$, $P_{A}$ and $P_{A'}$ in the expression for average success probability for the 2-bit POM game we get: $p_{Q}=\frac{1}{8}[4+\langle\Psi_{12}|\{A\otimes B+A'\otimes B +A\otimes B'- A'\otimes B'\}|\Psi_{12}\rangle ]$, where $A=2P_{A}-I$, $A'=2P_{A'}-I$, $B=2P_{B}-I$, $B'=2P_{B'}-I$ (see \cite{Supple}-(A) for details). Since all four operators (observables) $\{A,A',B, B'\}$ have eigenvalues $\{\pm 1\}\in [-1,1]$, and any operator from the set $\{A,A'\}$ commute with any operator from the set $\{B,B'\}$, by applying Cirel'son's result \cite{Cirelson} it follows that $\langle\Psi_{12}|\{A\otimes B+A'\otimes B+A\otimes B'- A'\otimes B'\}|\Psi_{12}\rangle \leq 2\sqrt{2}$. This gives, $p_{Q}\leq \frac{1}{2}[1+\frac{1}{\sqrt{2}}] $. We have already discussed that there exists quantum protocol to achieve this upper bound.~~$\blacksquare$

\emph{(b) Box world}: Interestingly, one can exceed the optimal quantum bound in the box world. This system can be understood as a black box taking a binary input $x = 0, 1$ and returning a binary output $a = 0,1$ \cite{Janotta}.
\begin{figure}[h!]
	\centering
	\includegraphics[height=4cm,width=8cm]{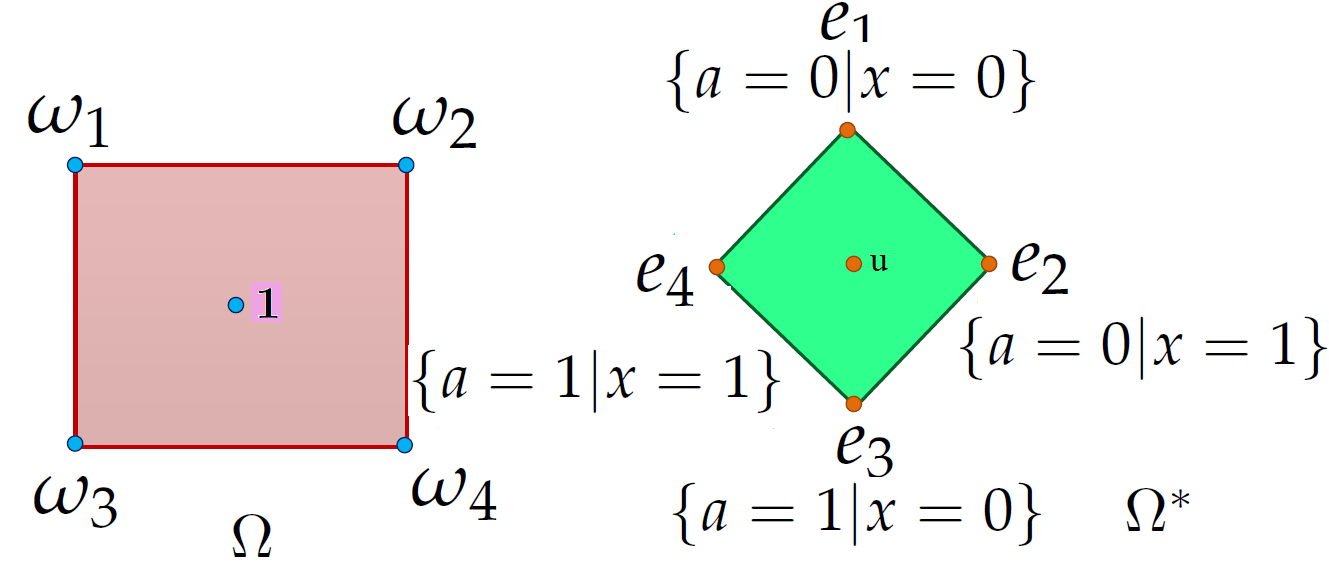}
	\caption{(Color on-line) $\Omega$ denotes the normalized state space for the Box world. Four corners denote four deterministic states and the central dot denotes the completely mixed state. $\Omega^*$ denotes the space of effects for the box world. $\{e_i|i=1,..,4\}$ are four extremal effects of the Box world. Two ideal measurements are $e_1+e_3=\mathbf{u}=e_2+e_4$.}\label{fig1}
\end{figure} 
The state of the system is described by a conditional probability distribution $P(a|x)$. The normalized state space $\Omega$ of the system can be represented as a square in $\mathbf{R}^2$ (see Fig.\ref{fig1}). The system thus features four pure states $\{\omega_j|j=1,..,4\}$. For each pure state, the outcome $`a$' is a deterministic function of the input $`x$' ($\omega_1\rightarrow a=0,~\omega_2\rightarrow a=x,~\omega_3\rightarrow a=1,$ and $\omega_4\rightarrow a=x\oplus 1$). The center of $\Omega$ is the maximally mixed state, that is, where $`a$' is independent of $`x$' and random. This maximally mixed state has non unique decomposition in term of pure states, i.e, $\frac{1}{2}\omega_1+\frac{1}{2}\omega_3=\frac{1}{2}\omega_2+\frac{1}{2}\omega_4=\mathbf{1}$. The space of effects, $\Omega^*$, is the dual of $\Omega$ (see Fig.\ref{fig1}). It features four extremal effects $\{e_j|j=1,..,4\}$ which correspond to the four measurement outcomes, i.e. obtaining output $`a$' for a given input $`x$'. The probability of $e_j$ on any state is easily determined. For instance, effect $e_1$ has probability one for states $\omega_{1},\omega_{2}$ and probability zero for states $\omega_{3},\omega_{4}$. There are two pure measurements for this system: the first is composed of effects $e_1$ and $e_3$, and corresponding to input $x = 0$; the second is composed of effects $e_2$ and $e_4$, and corresponding to input $x = 1$. Note that $e_1 +e_3 = e_2 +e_4 = \mathbf{u}$, where $\mathbf{u}$ is the unit effect. This system is also known as generalized bit (g-bit) \cite{Barrett}.

For performing the $2$-bit POM with better than quantum success Alice and Bob pursue the following strategy in the box world. Alice encodes her strings as:
\begin{eqnarray}
00\rightarrow \omega_1,~~~~11\rightarrow \omega_3,\nonumber\\
01\rightarrow \omega_2,~~~~10\rightarrow \omega_4,
\end{eqnarray}
and sends the encoded g-bit to Bob. The parity oblivious condition is satisfied as $\frac{1}{2}\omega_1+\frac{1}{2}\omega_3=\frac{1}{2}\omega_2+\frac{1}{2}\omega_4=\mathbf{1}$. To decode Alice's message, Bob performs (i) measurement $\{e_1,e_3|e_1+e_3=u\}$ if he wishes to learn the first bit and (ii) measurement $\{e_2,e_4|e_2+e_4=u\}$ if he wishes to learn the second bit
and he guesses Alice's bit as the measurement result. In every case, the guessed value is correct with certainty resulting $p_{box}(\beta_y=x_y)=1>\frac{1}{2}(1+\frac{1}{\sqrt{2}})=p^{opt}_Q(\beta_y=x_y)$. Clearly $p^{opt}_{box}(\beta_y=x_y)=1$ as the said strategy achieves 100\% success probability. 

\emph{(2) Correlation assisted scheme}: Let Alice and Bob now follow a different scheme in which instead of sending encoded states they use correlations of pre-shared bipartite states allowed in an operational theory. In an operational theory a general bipartite correlation can be thought as a probability distribution $p(\mathcal{C},\mathcal{D}|\mathcal{U},\mathcal{V})\equiv \{p(c,d|u,v)\}$, where $u\in\mathcal{U},~v\in\mathcal{V}$ are inputs given to Alice and Bob respectively and $c\in\mathcal{C},~d\in\mathcal{D}$ denote their respective outcomes. \emph{No-signaling} correlations satisfy the conditions $p(c|u)=\sum_{d\in\mathcal{D}}p(c,d|u,v);~\forall~c,u,v$ and the vice-verse. If local outcomes depend only on the choice of local measurements and (possibly) on the value of some shared (hidden) variable $\lambda\in\Lambda$ which takes values according to some distribution $p(\Lambda)=\{p(\lambda)\}$ then the correlation is called \emph{local}, i.e., $p_L(c,d|u,v)=\sum_{\lambda\in\Lambda}p(\lambda)p(c|u,\lambda)p(d|v,\lambda)$ for all $c,d,u,v$. Correlations which can not be expressed in such form are called \emph{nonlocal} \cite{Bell}. Entangled quantum particles \cite{Horodecki} exhibit nonlocal properties whereas they satisfy the no signaling conditions. 

Using bipartite correlations Alice and Bob can perform the POM task in the following manner. Alice prior to start of the POM game, shares a correlated pair of particles, prepared in the state (preparation) $P_{AB}\in\Omega_{AB}$, with Bob. Depending on the input string, given to her, Alice performs measurement on her particle of the correlated pair and sends the measurement result to Bob via classical communications (CC). Bob, receiving CC from Alice, performs operations on his particle and tries to guess Alice bit. However the CC should not contain any information about the parity of the Alice's input string. It turns out that local correlations are not useful for performing the POM task. 

{\bf Proposition-$2$}: \emph{Correlations having local description when assisted with classical communications are not useful for performing the parity-oblivious multiplexing task.} 

\emph{Proof}: Any local correlation between Alice and Bob can be thought as shared random variable $\lambda\in\Lambda$ taking values according to a probability distribution $p(\Lambda)$. In Ref.\cite{Spek09}, it has been shown that the only classical encodings of $x$ that reveal no information about any parity (while encoding some information about $x$) are those that encode only a single bit $x_i$ for some $i$. For simplicity, without loss of generality, consider that the shared variable takes discrete values $\lambda_i$ with $\sum_{i=1}^np(\lambda_i)=1$. If the variable takes value $\lambda_k$ then Alice encodes her $k^{th}$ bit and sends it to Bob. Bob, if asked, can correctly reveals this $k^{th}$ bit while for him other bits ($i\ne k$) are completely random. Thus using local correlation (shared randomness) Alice and Bob can design a strategy for determining only one bit with certainty. But, whichever bit Bob guesses correctly, the (average) success probability of Bob's guess is bounded by the noncontextual bound $\frac{1}{2}(1+\frac{1}{n})$. Therefore, due to convexity of distribution $p(\Lambda)$, it follows that any local correlation can not beat the noncontextual bound.~~$\blacksquare$ 

\emph{Remark}: Here it is important to note that in a POM game to obtain greater success than the classical bound, the theory need not contain nonlocal correlations. For example, consider a theory in which individual state space is identical as quantum state space but the composite state space is severely restricted than quantum state space. The state space of the composite system is \emph{minimal tensor product} \cite{Barnum} of individual Hilbert space and hence contains only separable states and hence no nonlocal correlation. In such a theory one can obtain the success probability of POM game as much as quantum theory by following the optimal encoding-decoding scheme of quantum theory. What the Proposition-2 proves is that if one wants to play the POM game by using correlation of such local theory she/he will not get any advantage.    

{\bf Proposition-$3$}: \emph{Any no-signaling correlation $\{p(ab|xy): a,b,x,y \in \{0,1\}\}$ violating Bell-CHSH inequality can exceed the classical bound for performing the 2-bit parity-oblivious multiplexing task. Moreover, if using a correlation average success in 2-bit POM game exceeds the quantum limit, then nonlocality of such correlation must exceed the Cirel'son bound.}

 \emph{Outline of Proof}: A proof follows from : (i) using the protocol for 2-bit random access code scenario discussed by Pawlowaski \emph{et al.} \cite{Pawlowski} in the context of information causality, and (ii) showing that this protocol respect the parity obliviousness condition. Then this implies that, any nonlocal correlation can achieve more than the classical limit for 2-bit POM task. Moreover, the quantum limit for 2-bit POM game nonlocality of correlation is restricted by the Cirel'son bound. We give a complete proof of the proposition in the supplementary \cite{Supple}-(C).~~$\blacksquare$

In quantum world using correlations of entangled particles Alice and Bob can win the POM game with more than classical (noncontextual) success probability. Using the steerability \cite{Schrodinger,Wiseman} of the entangled state and the classical communication Alice tries to prepare Bob's state in different preparations depending on the input string given to her. For achieving the best result Alice attempts to prepare Bob's particle into states which achieve optimal success probability in the encoding-decoding scheme. Sharing two-qubit maximally entangled state Alice can prepare the optimal states by echoing an identical procedure as in the remote state preparation protocol \cite{Bennett} (see \cite{Supple}-(B) for the protocol). 

However, presence of steering, alone, in a theory is not sufficient for achieving more than classical success probability; the theory must also be preparation contextual. For instance, there exists hypothetical \emph{toy bit} theory \cite{Spek07} which allows steering, but the success probability of POM in this theory is restricted to the classical bound as the theory is preparation noncontextual (see \cite{Supple}-(D)). On the other hand, though steerability in quantum theory is maximal, the optimal success probability of POM task is restricted due to its limited preparation contextuality.

To conclude, in this work we have considered an operational way to quantify the preparation contextuality of a general theory. We, then, show that quantum theory turns out to be less preparation contextual than another operational theory, namely box world. Further, we have shown that, in the quantum world, the restricted Bell-CHSH violation follows from limited preparation contextuality of the theory. Many researchers have tried to explain the limits of nonlocal feature in QM starting from a number of physically motivated ideas or principles. In particular, by considering various approaches, it has been successfully explained why Bell-CHSH quantity in quantum theory is restricted to Cirel'son bound \cite{Pawlowski}. Having established a link between the concept of nonlocality and preparation contextuality it would be interesting to suggest physical principal(s) leading to quantum bound on preparation contextuality.  

Recently, the authors of \cite{Chailloux} have shown that even-parity-oblivious encodings are equivalent to the INDEX game, which implies $2\rightarrow 1$ POM game is equivalent to the well known Bell-CHSH nonlocal game. Therefore, a quantum encoding of $2\rightarrow 1$ POM with average success probability $p_Q$ exists only if a quantum strategy for playing the Bell-CHSH game with the same average success probability exists. We take a different approach, by maximizing over all possible encoding-decoding schemes allowed in QM, we find the optimal success probability of $2$-bit POM game; it turns out to be restricted compared to a more general operational theory. We conclude that restricted preparation contextuality, therefore, bounds the winning probability of Bell-CHSH game (nonlocality) in quantum theory.        
    
\emph{Acknowledgments}: Thanks to Guruprasad Kar for many stimulating discussions. MB like to acknowledge useful conversations with Rajjak Gazi.  AM thanks Council of Scientific and Industrial Research, India for financial support through Senior Research Fellowship (Grant No. 09/093(0148)/2012-EMR-I). AsR likes to thank Andreas Winter for helping to prove the quantum optimality of 2-bit POM games. AA and AsR acknowledge support by the European Union Seventh Framework Programme (FP7/2007-2013) under the RAQUEL (Grant Agreement No. 323970) project, QALGO (Grant Agreement No. 600700) project, and the ERC Advanced Grant MQC.

\begin{widetext}
\section*{Supplementary}
\section*{(A):- Optimality of $2$-bit POM in quantum theory}

Alice prepares and encodes as $\{ij\longmapsto \rho_{ij}: i,j\in \{0,1\} \}$, where $\rho_{ij}$ are state operators acting on $\mathbb{C}^d$; she can always find an appropriate pure state $|\Psi_{12}\rangle \in \mathbb{C}^d\otimes \mathbb{C}^d$ and projectors $P_{A},P_{A'}$ such that:
\begin{eqnarray}\nonumber
	\frac{1}{2}\rho_{00}&=&\mbox{tr}_1 \{(P_{A}\otimes I)|\Psi_{12}\rangle\}, \nonumber\\
	\frac{1}{2}\rho_{11}&=&\mbox{tr}_1 \{((I-P_{A})\otimes I)|\Psi_{12}\rangle\},\nonumber \\
	\frac{1}{2}\rho_{01}&=&\mbox{tr}_1 \{(P_{A'}\otimes I)|\Psi_{12}\rangle\},\nonumber \\
	\frac{1}{2}\rho_{10}&=&\mbox{tr}_1 \{((I-P_{A'})\otimes I)|\Psi_{12}\rangle\}.\nonumber 
\end{eqnarray}	
 Alice performs one of the following projective measurements:
\begin{itemize}
\item[(i)] $P_{A}\otimes I + (I-P_{A})\otimes I =I\otimes I$, or
\item[(ii)] $P_{A'}\otimes I + (I-P_{A'})\otimes I =I\otimes I$
\end{itemize}
on part-$1$ of $|\Psi_{12}\rangle$ and depending on the measurement result she sends part-$2$ to Bob (or discard it). Measurement (i) and (ii) respectively produces two decompositions $\frac{1}{2}\rho_{00}+\frac{1}{2}\rho_{11}$ and $\frac{1}{2}\rho_{01}+\frac{1}{2}\rho_{10}$ for part-$2$ of $|\Psi_{12}\rangle$. Alice in this way prepares and send $\rho_{ij}$ to Bob; assured that parity obliviousness condition is always satisfied.
 
Bob on receiving part-$2$, performs a two outcome projective measurements $\{P_{B},(I-P_{B})\}$ ($\{P_{B'},(I-P_{B'})\}$), if he is asked to guess Alice's first (second) bit, and answers $0(1)$ when measurement outcome is $+1(-1)$. Here $P_{A},P_{A'},P_{B},P_{B'}$ are projectors acting on $\mathbb{C}^d$. Due to no constraint on the dimension of the Hilbert space, Neumark’s theorem allows us to consider only projective measurements, without loss of generality. The expression for average success probability for the 2-bit POM game thus become:
\begin{eqnarray}
p_{Q}&=& \frac{1}{8}\left[ \mbox{Tr}\left\lbrace \rho_{00}(P_{B}+P_{B'})\right\rbrace +\mbox{Tr}\{\rho_{01}(P_{B}+(I-P_{B'}))\}\right. \nonumber+\mbox{Tr}\{\rho_{10}((I-P_{B})+P_{B'})\}\nonumber \\
&&\left. ~~~~~~~~~~~~~~~~~~~+\mbox{Tr}\{\rho_{11}((I-P_{B})+(I-P_{B'}))\}\right] \nonumber \\
&=&\frac{1}{8}[4+ \mbox{Tr}\{(\rho_{00}-\rho_{11})P_{B}\}+\mbox{Tr}\{(\rho_{01}-\rho_{10})P_{B}\}\nonumber +\mbox{Tr}\{(\rho_{00}-\rho_{11})P_{B'}\}\nonumber \\
&&~~~~~~~~~~~~~~~~~~~~~~~~-\mbox{Tr}\{(\rho_{01}-\rho_{10})P_{B'}\}],\nonumber \\
&=&\frac{1}{8}[4+ 2\{\langle\Psi_{12}|(2P_{A}-I)\otimes P_{B}|\Psi_{12}\rangle \nonumber +\langle\Psi_{12}|(2P_{A'}-I)\otimes P_{B}|\Psi_{12}\rangle \\
&&~~~~~~~~~~~~+\langle\Psi_{12}|(2P_{A}-I)\otimes P_{B'}|\Psi_{12}\rangle-\langle\Psi_{12}|(2P_{A'}-I)\otimes P_{B'}|\Psi_{12}\rangle \}], \nonumber\\
&=& \frac{1}{8}[4+\langle\Psi_{12}|\{(2P_{A}-I)\otimes (2P_{B}-I)+(2P_{A'}-I)\otimes (2P_{B}-I)\nonumber\\ &&~~~~~~+(2P_{A}-I)\otimes (2P_{B'}-I)-(2P_{A'}-I)\otimes (2P_{B'}-I)\}|\Psi_{12}\rangle ], \nonumber \\
&=& \frac{1}{8}[4+\langle\Psi_{12}|\{A\otimes B+A'\otimes B+A\otimes B'- A'\otimes B'\}|\Psi_{12}\rangle ], \nonumber
\end{eqnarray}
where $A=2P_{A}-I$, $A'=2P_{A'}-I$, $B=2P_{B}-I$, $B'=2P_{B'}-I$. 

\section*{(B):- Entanglement assisted protocol in quantum mechanics}
Using correlation of entangled quantum particles, assisted with classical communication, Alice and Bob can win the POM game with better than classical success probability while satisfying the parity oblivious condition.
\begin{figure}[h!]
	\centering
	\includegraphics[height=4.5cm,width=5cm]{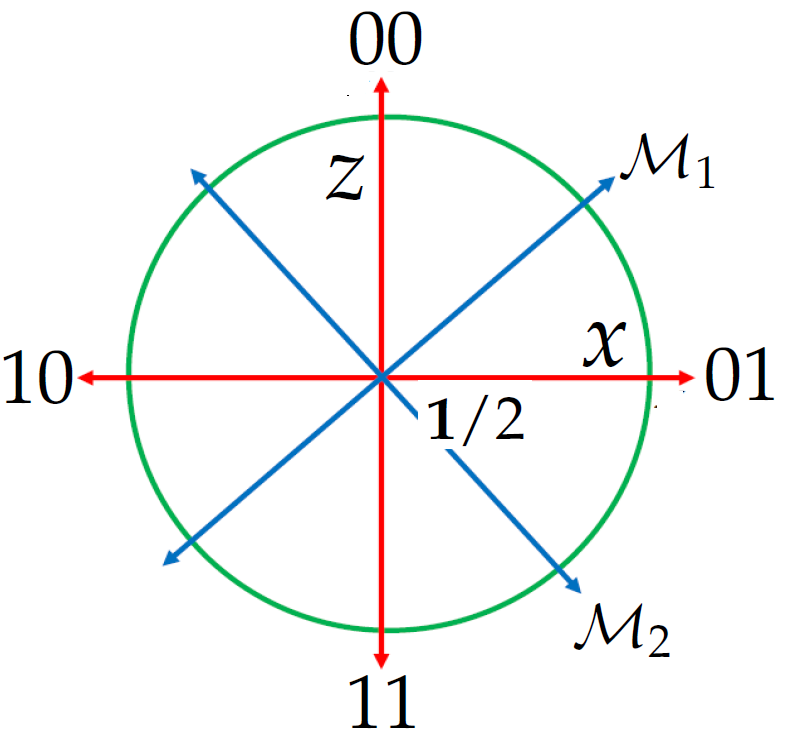}
	\caption{(Color on-line) The optimal quantum encoding-decoding scheme for playing the $2$-bit POM. Green circle denotes the equatorial circle on $X-Z$ plane of the Bloch sphere. Red and blue lines denote encoding and decoding scheme respectively.}\label{fig1}
\end{figure}

For achieving the optimal success Alice, depending on her input string, will try to prepare Bob's particle in the states in which she had encoded her strings in the optimal \emph{encoding-decoding} scheme (see Fig.\ref{fig1}), i.e., she tries to prepares Bob's particles accordingly:
\begin{eqnarray*}
00\rightarrow \frac{1}{2}(\mathbf{1}+\sigma_z),~~~11\rightarrow \frac{1}{2}(\mathbf{1}-\sigma_z),\\
01\rightarrow \frac{1}{2}(\mathbf{1}+\sigma_x),~~~10\rightarrow \frac{1}{2}(\mathbf{1}-\sigma_x).
\end{eqnarray*}

Let Alice shares a correlated pair of particles with Bob prepared in singlet state $|\psi^-\rangle_{AB}=\frac{1}{\sqrt{2}}(|0\rangle_A\otimes |1\rangle_B-|1\rangle_A\otimes |0\rangle_B)$. If the input sting is $00$ or $11$ then Alice performs $\sigma_z$ measurement on her part of singlet pair otherwise she performs $\sigma_x$ measurement. In both cases the \emph{unconditional} state of Bob's particle is $\frac{1}{2}\mathbf{1}$. Hence by performing most general quantum measurement Bob can not get any information about the parity of the input string.

Whenever Alice's input string is $00$ she performs $\sigma_z$ measurement on her particle. If her measurement outcome is $`1$' then due to anti-correlation of singlet correlation Bob's particle is prepared in the state $\frac{1}{2}(\mathbf{1}+\sigma_z)$. But, if her measurement result is $`0$' then state of the Bob's particle is $\frac{1}{2}(\mathbf{1}-\sigma_z)$ and it requires Bob to perform a rotation on his particle so that the state of the particle ends is the desired state $\frac{1}{2}(\mathbf{1}+\sigma_z)$. So, depending on her measurement result, Alice sends classical communication to Bob to inform whether he needs to perform the rotation or not. Note that this classical communication does not carry any information about the parity of the string. In rest of the cases Alice follows similar procedures and succeeds in preparing Bob's particle in the desired states. Note that this protocol echoes the same strategy as followed in the remote state preparation protocol \cite{Bennett}.   

For guessing Alice's bit, Bob performs the measurements as done done in the optimal \emph{encoding-decoding} scheme. For decoding the first bit he performs the spin measurement $(\sigma_z+\sigma_x)/\sqrt{2}$ and for second bit he performs spin measurement $(\sigma_z-\sigma_x)/\sqrt{2}$ and guesses the bit depending on the measurement result, which results in the $\frac{1}{2}(1+\frac{1}{\sqrt{2}})$ winning probability of the POM task.	

\section*{(C):- Complete proof of the Proposition-3 in the main text}
{\bf Proposition-$3$}: \emph{Any no-signaling correlation $\{p(ab|xy): a,b,x,y \in \{0,1\}\}$ violating Bell-CHSH inequality can exceed the classical bound for performing the 2-bit parity-oblivious multiplexing task. Moreover, if using a correlation average success in 2-bit POM game exceeds the quantum limit, then nonlocality of such correlation must exceed the Cirel'son bound.}

\emph{Proof:} We start with a definition for nonlocality of correlation $p(ab|xy)$. The Bell-CHSH inequality expressed in terms of these correlations take a form: $\mathbb{B}= \sum^{1}_{x=0}\sum^{1}_{y=0} p(a\oplus b=xy|x,y) \leq 3$. Any correlation for which $\mathbb{B}>3$ is nonlocal, maximum nonlocality achievable in quantum mechanics is Cirel'son bound $2+\sqrt{2}$. Over set of all no-signalling correlation $\mathbb{B}$ can achieve the maximum value $4$. By a suitable local randomization technique used in \cite{masanes}, keeping nonlocality parameter $\mathbb{B}$ invariant, one can write such no-signaling correlation in a canonical form: $p(a\oplus b=xy|x,y) =\frac{1}{2} (1+ \gamma)$; where $0\leq \gamma \leq 1$. For $\gamma>1/2$ correlations are nonlocal and $\gamma=1/\sqrt{2}$ correspond to Cirel'son bound $\mathbb{B}_{Q}=2+\sqrt{2}$. 

Suppose, using canonical correlations, Alice and Bob play the $2$-bit POM game with the same protocol as in 2-bit random access code scenario discussed by Pawlowaski \emph{et al.} \cite{Pawlowski} in the context of information causality. Let us denote Alice's two bit string as $x_1x_2$, where $x_1,x_2\in\{0,1\}$. Alice inputs $x=x_1\oplus x_2$ and obtains $a$ as outcome. Alice communicates $c=x_1\oplus a$ to Bob. If Bob is asked to guess Alice's first bit he gives $y=0$ as input otherwise he inputs $y=1$. After obtaining outcome $b$ from the shared canonical correlation and communication $c$ from Alice he declares his answer as $b\oplus c$. With this protocol one can check that the average success probability of correct guess by Bob is $\frac{1}{2} (1+ \gamma)$. Note that the parity obliviousness condition is satisfied in this protocol due to: i) correlation used, being a no-signaling resource, can not carry any information, and ii) the communicated bit $c$ contains no information about the parity of Alice's string. Now it is easy to see that, with this protocol, any nonlocal correlation achieves more than the classical limit for 2-bit POM task. Moreover, the quantum limit for 2-bit POM game implies that $\gamma\leq 1/\sqrt{2}$ which can hold only if $\mathbb{B}_{Q}\leq 2+\sqrt{2}$, i.e., nonlocality of correlation is restricted by the Cirel'son bound. ~~$\blacksquare$  

\section*{(D):- Steerable toy-bit theory is not useful for POM}
Spekkens has introduced a toy theory in order to argue for an epistemic view of quantum states \cite{Spek07}. The theory is based on a principle, namely \emph{knowledge balance principle}, according to which the number of questions about the physical state of a system that are answered must always be equal to the number that are unanswered in a state of maximal knowledge.

For the elementary system the number of questions in the canonical set is two, and consequently the number of \emph{ontic} states is four. Denote the four ontic states
as $`1$',$`2$',$`3$', and $`4$'. An \emph{epistemic} state is nothing but a probability distribution $\{(p_1,p_2,p_3,p_4)|~p_i\ge0~\forall i~\&~\sum_{i=1}^{4}p_i=1\}$, over the ontic states. Denoting disjunction by the symbol $`\vee$'(read as \emph{or}), the six possible pure \emph{epistemic} states, allowed by the knowledge balance principle, read as:
\begin{eqnarray*}
1\vee 2\leftrightarrow \{\frac{1}{2},\frac{1}{2},0,0\}=\includegraphics[height=.5cm,width=2cm]{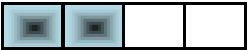},\\
3\vee 4\leftrightarrow \{0,0,\frac{1}{2},\frac{1}{2}\}=\includegraphics[height=.5cm,width=2cm]{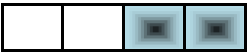},\\
1\vee 3\leftrightarrow \{\frac{1}{2},0,\frac{1}{2},0\}=\includegraphics[height=.5cm,width=2cm]{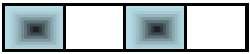},\\
2\vee 4\leftrightarrow \{0,\frac{1}{2},0,\frac{1}{2}\}=\includegraphics[height=.5cm,width=2cm]{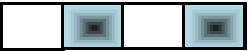},\\
1\vee 4\leftrightarrow \{\frac{1}{2},0,0,\frac{1}{2}\}=\includegraphics[height=.5cm,width=2cm]{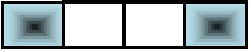},\\
2\vee 3\leftrightarrow \{0,\frac{1}{2},\frac{1}{2},0\}=\includegraphics[height=.5cm,width=2cm]{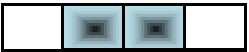}.
\end{eqnarray*}

For such a system, one have less than maximal knowledge if both questions in the canonical set to be unanswered. This corresponds to the epistemic mixed state:
\begin{equation*}
1\vee 2\vee 3\vee 4\leftrightarrow \{\frac{1}{4},\frac{1}{4},\frac{1}{4},\frac{1}{4}\}=\includegraphics[height=.5cm,width=2cm]{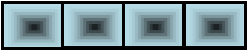}.
\end{equation*}
The mixed state has following different convex decompositions:
\begin{eqnarray}\label{decom1}
1\vee 2\vee 3\vee 4 &=& (1\vee 2)+_{cx}(3\vee 4),\\\label{decom2}
	&=& (1\vee 3)+_{cx}(2\vee 4),\\\label{decom3}
	&=& (1\vee 4)+_{cx}(2\vee 3).
\end{eqnarray}
The above set of decompositions can be thought as different preparation procedures for the same mixed state, a phenomenon that can be observed in quantum theory. 

Compatible with knowledge balance principle, there are two types of (pure) epistemic states for a pair of elementary systems:
\begin{itemize}
\item[(1)] $(a\vee b).(c\vee d)\equiv (a.c)\vee (a.d)\vee (b.c)\vee (b.d)$; where $a,b,c,d\in{1,2,3,4}$ and $a\ne b,~c\ne d$.
\item[(2)] $(a.e)\vee (b.f)\vee (c.g)\vee (d.f)$; where $a,b,c,d,e,f,g,h\in\{1,2,3,4\}$ and $a, b, c, d$ are all different and same is for $e, f, g, h$.
\end{itemize}
For the second type of states the state for marginal elementary systems (both) is $1\vee 2\vee 3\vee 4$. Let Alice shares with Bob a bipartite elementary system prepared in the state $(1.1)\vee (2.2)\vee (3.3)\vee (4.4)$. If Alice implements the measurement that distinguishes $1\vee 2$ from $3\vee 4$ on her part of the system then will be able to remotely prepare Bob system in decomposition of Eq.(\ref{decom1}). Similarly implementing measurements that distinguishes $1\vee 3$ from $2\vee 4$ and $1\vee 4$ from $2\vee 3$ she can prepares other two decompositions, i.e., decomposition of Eq.(\ref{decom2}) and Eq.(\ref{decom3}) respectively, which establishes steering like phenomena for toy-bit theory.

A model is said to be preparation non-contextual if the probability distributions over the ontic states corresponding to different preparation procedures of an identical operational state remains identical. Otherwise the model will be called preparation contextual. It has been proved that any mixed quantum state is preparation contextual \cite{Banik,Spek05}. But in this case, though the mixed state $(1\vee 2\vee 3\vee 4)$ can be prepared in different ways (decompositions (\ref{decom1}), (\ref{decom2}) and (\ref{decom3})) all the preparations give identical probability distribution $\{\frac{1}{4},\frac{1}{4},\frac{1}{4},\frac{1}{4}\}$ over the ontic states $\{1,2,3,4\}$ for every preparation procedure. Therefore the theory is preparation non-contextual. Hence though the toy-bit theory exhibits steering phenomena, but being preparation noncontextual it gives no advantage in POM task.
\end{widetext}

\end{document}